\documentclass[10pt]{article2}
\usepackage{latexsym, epsfig}
\usepackage{amsmath}
\usepackage{amsfonts}
\usepackage{url}
\input{Qcircuit}

\newcommand\beq{\begin{equation}}
\newcommand\eeq{\end{equation}}
\newcommand\bea{\begin{eqnarray}}
\newcommand\eea{\end{eqnarray}}

 \def\squarebox#1{\hbox
to #1{\hfill\vbox to #1{\vfill}}}
\def\qed{\hspace*{\fill}\vbox{\hrule\hbox{\vrule\squarebox{.667em}\vrule}\hrule}
} 

\newcommand{\ba}{\begin{array}}
\newcommand{\ea}{\end{array}}

\newtheorem{theo}{Theorem}

\newtheorem{defi}{Definition}

\newtheorem{lem}{Lemma}

\begin{document}

\title{Fault-Tolerant Quantum Computation for Local Leakage Faults}
\author{Panos Aliferis\thanks{Institute for Quantum Information, Caltech, Pasadena, CA 91125, USA.
\texttt{panos@theory.caltech.edu} } \hspace{0.05cm} and Barbara M. Terhal\thanks{IBM
Watson Research Center, Yorktown Heights, NY 10598, USA.
\texttt{terhal@watson.ibm.com} }}
\date{\today}

\maketitle

\begin{abstract}
We provide a rigorous analysis of fault-tolerant quantum computation
in the presence of local leakage faults. We show that one can
systematically deal with leakage by using appropriate
leakage-reduction units such as quantum teleportation. The leakage
noise is described by a Hamiltonian and the noise is treated
coherently, similar to general non-Markovian noise analyzed in Refs.
\cite{Terhal04} and \cite{Aliferis05b}. We describe ways to limit
the use of leakage-reduction units while keeping the quantum
circuits fault-tolerant and we also discuss how leakage reduction by
teleportation is naturally achieved in measurement-based
computation.
\end{abstract}

\section{Introduction}

There has been much recent work in the theory of fault-tolerant
quantum computation. The foundations of this theory were initially
developed in Refs.$\,$\cite{Shor96}, \cite{Kitaev97},
\cite{Aharonov99comb}, \cite{Knill96b}, \cite{Gottesman97} and
\cite{Preskill97}. The more recent work on fault-tolerance builds
on, improves and extends these fundamental results in various
directions. First of all, there is research on finding good code
architectures and determining the corresponding threshold noise
values and overheads (e.g., see \cite{Steane99, Steane02}). Current
threshold estimates for the [[7,1,3]] code range from $O(10^{-3})$
\cite{Steane02, Reichardt04} to $O(10^{-4})$ \cite{Svore04} with a
$O(10^{-5})$ lower bound established in Ref.$\,$\cite{Aliferis05b}.
Recent work by Knill \cite{Knill04} shows that using small detection
codes combined with extensive gate teleportation can lead to a
threshold of $O(10^{-2})$, albeit at a large overhead in resources.

Secondly, there is a growing body of work on estimating noise
threshold values taking into account the spatial architecture and
means for qubit transportation \cite{Svore04, Metodi05} or
measurement times \cite{Steane02}. Analyses pertaining to actual
physical systems, such as ion traps \cite{Steane04} or optical
quantum computers \cite{Nielsen05} are starting to be developed.

Thirdly, there has been research extending the error models and
methods used in the theory of quantum fault-tolerance. In Ref.$\,$\cite{Terhal04} a first fault-tolerance analysis was carried out for local non-Markovian noise. This result was subsequently generalized
in Ref.$\,$\cite{Aliferis05b}. In the last paper the foundations of
quantum fault-tolerance were strengthened and a rigorous proof of
the threshold theorem for distance-3 codes was given.

In this paper we follow this third line of research. We will put
some previously scattered results and ideas on a rigorous footing
and formulate a fault-tolerance analysis for leakage faults. In many
physical scenarios a two-level qubit is a subspace of a
higher-dimensional space. This can happen when we use encoded
physical states as elementary qubits as in the
decoherence-free-subspace (DFS) formalism. More typically, the qubit
is simply two low-lying levels in a many-level physical system. Some
examples are ion trap qubits, optical qubits based on the KLM scheme
\cite{Knill00} where the qubit is the dual-rail subspace of two
photonic modes, Josephson junctions where the qubit is formed by the
lowest energy states in a double well potential \cite{Orlando99},
and encoded electron-spin qubits in quantum dots \cite{DiVincenzo00,
Petta05}. For such qubits, leakage faults that transfer amplitude in
and out of the qubit-subspace are likely to be an important source
of errors. In this paper we will refer to the qubit-subspace that is
part of a larger extended space as the {\em system-space}.

As with regular faults, the first protection against leakage faults
is the use of `low decoherence' qubit encodings and dynamical
decoupling methods. The method of dynamically decoupling leakage
faults by sequences of pulses has been explored in
Refs.$\,$\cite{Tian00,Byrd05}. In this paper we will analyze how the
(remaining) leakage faults can be dealt with by means of
error-correction. The problem is that error-correction is not
designed to deal with leakage faults directly. Thus we need to
convert leakage faults to regular faults, i.e. faults that occur in
the system-space and that can be corrected by error-correcting
codes. The reduction to regular faults can be of two kinds. One
could detect that leakage occurred (using e.g. the circuit
presented in \cite{Gottesman97,Preskill97}) and replace the leaked
qubit by a new physical qubit in the system-space. Alternatively, a
leakage fault can be reduced to a regular fault without us learning
whether the leakage occurred. We will generally refer to this second
tool as a Leakage-Reduction Unit, or LRU. Quantum teleportation is
the most natural implementation of a LRU \cite{Mochon04}.

In this paper we will focus on LRUs since they can be implemented
{\em universally} by teleportation. The ability to detect leakage on
the other hand depends on the specific implementation and the nature
of the leakage space. We will discuss the use of leakage detection
and how our results can be used in leakage-detecting implementations
in \S\ref{sect:discuss}. Of course in our analysis we assume that
LRUs (or the more powerful leakage-detection units) are subject to
regular and leakage faults themselves.

Let us sketch an overview of the results in this paper. A
fault-tolerant threshold analysis consists of two fairly independent
parts. First one needs to prove that if the computation is {\em
good} (i.e., has {\em few} faults), the logical dynamics of the
encoded computation is {\em correct} under some technical definition
of good, few and correct. This analysis for leakage faults will be
the essential contribution of this paper. Secondly, one needs to
show that the probability (or amplitude or norm depending on which
error model one uses) for a computation which is {\em bad} (i.e.,
has {\em many} faults) can become as small as desired below the noise threshold. Putting
these two things together then gives rise to the threshold theorem,
Theorem \ref{theo:thresh}. The second part depends on the error
model that describes leakage faults.
%
%
The treatment of leakage noise has been informally discussed in previous literature \cite{Knill96b,Preskill97}. However, physical leakage noise cannot be accurately captured by a simple Markovian noise model as in those works. In order to do a complete analysis of the effect of leakage faults, in \S\ref{sec:leakmodel} we
consider a microscopic Hamiltonian model in which leakage processes are treated
coherently. In this sense the leakage error model is very similar to
the general non-Markovian model which was analyzed in
Refs.$\,$\cite{Terhal04} and \cite{Aliferis05b}. We will use the
results in these papers directly to bound the norm of the sum of bad
fault operators in the encoded computation.

In the analysis of leakage faults given in \S\ref{sec:leakanalysis}
we assume that LRUs are placed before every elementary gate in the
encoded circuit. Then in \S\ref{sec:fewerLRU} we show that this frequent placement of
LRUs is unnecessary and we give conditions for when LRUs can be
omitted. In Appendix \ref{sec:example} we illustrate these concepts with circuits for the Steane [[7,1,3]] code.

Given a fault-tolerant circuit with LRUs before each gate, it is
clear that, {\em if} the LRUs work without error, this circuit
subjected to leakage faults is equivalent to a circuit without LRUs
subjected to regular faults. This is because LRUs replace any leaked
input with a state in the system-space before the next gate is
applied. However a LRU can fail. It can fail due to a regular fault, but it can also fail due to a
leakage fault. For example, the Bell state that is created for
quantum teleportation of leaked qubits can itself be leaky on the
outgoing qubit. In order to deal with leaky LRUs, we will set-up an
analysis that is very much like the rigorous analysis in Ref.$\,$\cite{Aliferis05b}. In particular, we
need to show that good fault-paths with few leakage or regular
faults give correct answers since errors are being corrected and do
not spread too badly. The essential insight is that LRUs can be
viewed as performing leakage-correction at the lowest unencoded
level.

In \S \ref{sec:meas-based} we discuss how implementing gates by
teleportation \cite{Gottesman99} provides an alternative method for
protecting against leakage faults. This is an approach which is
naturally embodied in measurement-based models of computation. We
focus in particular on circuit simulations in the graph-state model
\cite{Raussen03} and require that graph states of a certain
`standard' form are used. Such graph states are of potential
practical interest as they appear in several proposals relating to
quantum computation with non-deterministic gates (e.g., see
\cite{Nielsen05,Browne04, Lim05}). We would like to emphasize that our analysis for graph-states is suitable for
leakage or qubit loss that cannot be detected; for detectable qubit
loss that is error-free there are more efficient methods such as the
ones described in \cite{Varnava05}.

\subsection{Notation}
The standard convention for circuit diagrams is that time moves from
left to right. On the other hand, the representation of a circuit as
a sequence of mathematical operations applied to an input is usually
represented in reverse order, i.e. as $A_k \ldots A_2 A_1
\ket{\psi}$. We will occasionally also denote the operation $A_k
\ldots A_2 A_1$ as $A_1 \ast A_2 \ast \ldots \ast A_k$, i.e. with a
circuit diagram timing convention. We also use the notation
$\sigma_{\rm x}$, $\sigma_{\rm y}$ and $\sigma_{\rm z}$ for the
usual Pauli operators.

\section{The Leakage Error Model}
\label{sec:leakmodel}

We begin by specifying our leakage error model. In this paper we focus on leakage faults, but our analysis also permits mixed error models. We assume the following Hamiltonian
\beq H=H_{\rm ideal}(t)+H_{\rm faults}(t), \eeq
\noindent where $H_{\rm faults}(t)=H_{\rm regular}(t)+H_{\rm
leak}(t)$. $H_{\rm ideal}(t)$ generates the ideal system dynamics
implementing the encoded computation. $H_{\rm
regular}(t)=H_{S(B)}(t)+H_B(t)$ where $H_{S(B)}(t)$ represents a
local Hamiltonian coupling to a (non-Markovian) environment as in
Refs.$\,$\cite{Aliferis05b} and \cite{Terhal04} or faults not
involving a bath. As in these papers we assume that
\beq H_{S(B)}(t)=\sum_{\alpha} H_{S(B),\alpha(t)} \; \;,
\label{eq:local-decompbath} \eeq
\noindent where $H_{S(B),\alpha(t)}$ only acts on the set of qubits involved in a particular location\footnote{A location is an elementary operation in the fault-tolerant simulation, such as a qubit-preparation, a single-qubit measurement or a quantum gate.} $\alpha(t)$ in the ideal computation at time $t$ and potentially the bath or environment $B$.

Let ${\cal H}_S$ be the Hilbert space of the system qubits, with ${\cal H}_{S[i]}$ the Hilbert space of qubit $i$. There is an extension of each ${\cal H}_{S[i]}$, ${\cal H}_{S_{ext}[i]}={\cal H}_{S[i]} \oplus {\cal H}_{L[i]}$ so that ${\cal H}_{L[i]}$ is the leakage space of qubit $i$.
We assume that ${\cal H}_{L}=\otimes_i {\cal H}_{L[i]}$, i.e. the leakage spaces of the individual qubits are disjoint.
The leakage part of the Hamiltonian $H_{\rm leak}(t)$ is of the form
\beq H_{\rm leak}(t)=H_{SL(B)}(t)+H_L(t) \, , \eeq
\noindent where $H_{SL(B)}(t)$ is a linear combination of operators
coupling the leakage and system spaces and potentially some
environment $B$. An example of leakage which includes a coupling to an environment
is a trapped ion in a cavity where transitions into or out of
the two-level subspace can create or annihilate photons in the
cavity.
$H_L(t)$ describes the evolution in the leakage spaces and any
coupling of the leakage space to the environment.
We require that the coupling Hamiltonian is {\em local} meaning that
\beq H_{SL(B)}(t)=\sum_{\alpha} H_{SL(B),\alpha(t)} \; \;,
\label{eq:local-decomp} \eeq
\noindent where $H_{SL(B),\alpha(t)}$ only leaks from/to the
system/leakage space of the set of qubits involved in a particular
location $\alpha(t)$ in the ideal computation at time $t$.

We note that leakage can be an inherently non-Markovian process in
particular when it does not involve additional environments. Since
the leakage-space and the system-space form a direct sum and not a
direct product, there is no meaning to the notion of loss of
information by tracing out a subsystem.
Note that we restrict the leakage space to be local and different
for each qubit; this condition is fulfilled in most physical
systems.

The interaction structure of the bath and the interaction among the
leakage spaces in $H_{\rm faults}(t)$ can be of two kinds. In the
first one we assume that the leakage spaces of qubits $i$ and $j$
are disjoint and only interact during the time that qubits $i$ and
$j$ interact. For leakage or regular faults involving a bath, we
also assume each qubit $i$ has associated with it an environment
space ${\cal H}_{B[i]}$, with environment spaces also being disjoint
and only interacting when the corresponding qubits interact. This
model is basically the natural generalization of the non-Markovian
model considered in Ref.$\,$\cite{Terhal04} to leakage noise. In
this scenario every location $\alpha$, involving, say,  a single
qubit $i$, can be described by a unitary operator $U[i]$ acting on
the extended space of qubit $i$ and possibly its environment space,
i.e. $U[i]\colon {\cal H}_{S_{ext}[i]} \otimes {\cal H}_{B [i]}
\rightarrow {\cal H}_{S_{ext}[i]} \otimes {\cal H}_{B [i]}$.
We can write this operator as
\beq U[i]=U_0[i]+E[i] \, , \label{eq:eexpansion} \eeq
\noindent where $U_0[i] \colon {\cal H}_{S[i]}\otimes {\cal H}_{B
[i]}\rightarrow {\cal H}_{S[i]}\otimes {\cal H}_{B [i]}$ is the
ideal gate on the system and $E[i]$ is the fault operator. As in
Lemma 2 in \cite{Terhal04}, one can bound $||E|| \leq \epsilon
\equiv \epsilon_{\rm reg} + \epsilon_{\rm leak}$, where\footnote{Here $t_0$ is the time to execute an elementary operation and $|| \cdot ||$ is the sup norm (see \cite{Terhal04}). The factor $2$ comes from considering two-qubit gates.}
$\epsilon_{\rm reg}=2 t_0 \max_{\, \alpha} \, ||H_{S(B),\alpha}||$
and $\epsilon_{\rm leak}=2 t_0 \max_{\, \alpha} \,
||H_{SL(B),\alpha}||$. The amplitude $\epsilon$ is assumed to be small
and will enter the threshold theorem, Theorem \ref{theo:thresh}.

In the second case, we let the local leakage spaces interact in an
arbitrary manner and assume a common environment for all locations.
This is similar to the general non-Markovian model that was analyzed
in Ref. \cite{Aliferis05b}. In this case we cannot identify a
unitary operation per location but one can prove the essential {\em
local noise} Lemma 6 as in \cite{Terhal04}, namely

\begin{lem}[Local Noise Lemma]
Consider the entire unitary evolution $U$ of an (encoded) quantum
computation subject to local leakage or regular faults described by $H_{\rm faults}(t)$. Let $\epsilon_{\rm leak}$ be the leakage amplitude, and let
$\epsilon_{\rm reg}$ be the amplitude for regular faults as defined above. We expand
$U$ as a sum over fault-path operators which are characterized by a
set of faulty locations $\mathcal{I}$. A fault-path operator with
$k$ faults, denoted as ${\cal I}_k$, has norm bounded by
\beq
||E(\mathcal{I}_k)|| \leq  \epsilon^k \, , \hspace{0.2cm} {\rm with} \hspace{0.2cm} \epsilon \equiv \epsilon_{\rm leak} + \epsilon_{\rm reg} \, .
\eeq
\label{lem:local-noise}
\end{lem}

The fault-operator $E[i]$ of Eq. (\ref{eq:eexpansion}) or the
operators appearing in the fault-path expansion in
Lemma$\,$\ref{lem:local-noise} can be expanded in terms of a basis
of error operators. In our analysis we will use the notions of a
`regular' and a `leakage' error-operator, $R$ and $L$ respectively,
whose action on the extended space of a qubit has the form
\beq
R = \left( \begin{array}{cc} A & 0 \\ 0 & 0 \end{array} \right) \, , \, L= \left( \begin{array}{cc} 0 & C \\ B & 0 \end{array} \right) \, .
\eeq
\noindent Here $A$ is an operator acting on the system-space alone, and $B$, $C$ are operators coupling the system-space with the corresponding leakage-space.

We would like to add a comment concerning the identification of the error amplitude.
In our simplest error model the error-free evolution of a gate is of the form $U_0=(U_0^S \oplus U_0^{L}) \otimes U_0^B$.
In principle we could generalize this notion of error-free evolution and say that a gate is {\em leakage reducing
error-free} if the following holds: (1) if the input is contained in
the system-space the gate performs the ideal unitary gate in the
system-space and (2) for all inputs that are not contained in the
system-space, the leakage probability $|\beta|^2$ of the output
state $\ket{\psi_{\rm out}}=\alpha \ket{\psi_S}+\beta\ket{\psi_L}$
is smaller or equal that the leakage probability of the input state.
When subjected to a state partially in the leakage space, such a
gate would help reduce the leakage fault to a regular fault. An
example could be the physical process of decay of higher excited
states (in an ion, atom, quantum dot etc.) back to the lower-lying
system states. One has to be cautious in understanding the
nature of such process. In general an operation such a
\beq
\left( \begin{array}{cc} U_0^S & 0 \\ 0 & U_0^L \end{array} \right)+\left( \begin{array}{cc} 0 & C \\ 0 & 0 \end{array} \right),
\eeq
could both coherently reduce as well as amplify the leakage amplitude of any incoming state
$a \ket{\psi_S}+b\ket{\psi_L}$. The amplification of the leakage amplitude can come about by
negative interference between $U_0^S \ket{\psi_S}$ and $C \ket{\psi_L}$.
If such negative interference can be excluded, for example, the processes are incoherent and evolve
other environments which prevent interference, then these processes may be counted as part of the
error-free evolution. Thus, depending on the
particular implementation and modeling of physical processes involving leakage, it may
be possible to use a reduced error amplitude or probability, treating as beneficial those processes that
naturally reduce leakage faults to regular faults.

\section{Leakage Fault-Analysis}
\label{sec:leakanalysis}

Consider the fault-tolerant simulation of some ideal computation. By
a standard choice of a universal set of operations, a location in a
quantum circuit (0-Ga) is either an elementary single- or two-qubit
gate (including the identity gate which realizes a memory (wait) step), a single-qubit preparation or a
single-qubit measurement. To obtain the fault-tolerant simulation,
every 0-Ga in the ideal circuit is replaced in the level-$1$
simulating circuit by a rectangle (1-Rec) which consists of a
fault-tolerant encoded gate (1-Ga) {\em followed} by
error-correction ($1$-EC) on each encoded block. The preparation
1-Rec contains a fault-tolerant 1-preparation circuit followed by
1-EC, whereas the measurement 1-Rec contains only the fault-tolerant level-$1$ measurement.

Repeated application of the replacement rule gives rise to the level-$k$ simulation. In this way,
each location in the ideal circuit gets replaced by a $k$-rectangle ($k$-Rec) which consists of
a fault-tolerant level-$k$ gate ($k$-Ga) followed by error-correction at level $k$ ($k$-EC). For
the analysis of the level-$k$ circuit, Ref.$\,$\cite{Aliferis05b} defines an extended $k$-rectangle,
a $k$-exRec, as a $k$-rectangle grouped together with the $k$-ECs preceding it on all its inputs.

Before proceeding, let us first discuss the essential difference
between leakage and regular faults. The problem with leakage faults
that distinguishes them from regular qubit faults is that, once a
qubit has leaked, future interactions with other qubits, even if
ideal, can cause these other qubits to become erroneous as well. In
the presence of leakage, a 0-Ga acting on a leaked input will
operate on the extended space of its input qubits even if it is
executed ideally. Therefore, leakage errors cannot be propagated in
a simple way through ideal gates in the circuit since this
propagation depends on the particular gate implementation (i.e., the
specification of how gates operate on the extended space). As
fault-tolerant circuit design is guided by the particular ways in
which errors propagate through the circuit, in the presence of
leakage errors the design of the circuit will in general no longer
be effective in maintaining fault-tolerance.

To ensure that ideal 0-Ga's operate on the system-space of their
inputs we will make use of leakage-reduction units (LRUs) and place
a LRU before every \emph{gate} 0-Ga. We do not place LRUs preceding measurement 0-Ga's which achieve leakage reduction by themselves or preceding qubit-preparation 0-Ga's which have no input. LRUs, when executed without
faults themselves, guarantee that (a) if their input is in the
system-space then the identity operation is performed, and (b) if
their input in the leakage-space, then their output is some state in
the system-space. Clearly, teleportation is a natural way to satisfy
these two requirements.

LRUs can be viewed as performing `leakage correction' at level $0$
of the fault-tolerant simulation, just as $k$-EC gadgets perform
regular error-correction at level $k$. The strategy in our analysis
will be to prove the `LRU-reduction' Lemma 2 that will allow us to
transform the initial simulation where level-$0$ leakage correction
is performed by LRUs to an equivalent simulation where this level of
correction is removed and the effective faults are only regular
qubit faults. After this step, we can apply the threshold theorem
for regular faults to this equivalent simulation thus proving the
threshold theorem for our leakage error model.

We first need to define new notions of goodness and correctness at level 0. In the case of regular faults this would be trivial: a 0-Rec is just a 0-Ga and it is bad (and incorrect) when it is faulty. Recall that to deal with leakage faults we have placed LRUs on every qubit preceding every gate 0-Ga in the level-$k$ simulation. In analogy to 1-Recs, we can now define LRU-0-Recs (see Fig.$\,$1).
\begin{defi}[LRU 0-Rectangles]
For a gate {\rm 0-Ga}, a {\rm LRU-0-Rec} is the {\rm 0-Ga} preceded by {\rm LRUs} on all its inputs. For a qubit-preparation or measurement {\rm 0-Ga}, the {\rm LRU-0-Rec} coincides with the {\rm 0-Ga}.
\label{def:lru-rect}
\end{defi}
\begin{figure}[h]
\begin{center}
\epsfig{file=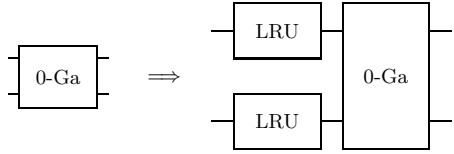,width=6cm} \label{lru-repl}
\caption{\footnotesize{To combat leakage faults, we insert LRUs on
all inputs preceding every gate 0-Ga in the level-$k$ circuit. The
combination of a 0-Ga with its preceding LRUs we call a LRU-0-Rec.}}
\end{center}
\end{figure}

After LRUs are inserted in the level-$k$ circuit, we can think of
1-exRecs as being transformed to what we can call LRU-1-exRecs.
Similarly, $k$-exRecs are transformed to LRU-$k$-exRecs, etc.
Inserting LRUs before 0-Ga's intends to convert existing leakage
faults to regular faults before the 0-Ga is applied.  However, LRUs
can be faulty themselves and this fact motivates the following
definition.
\begin{defi}[LRU Good, Bad, Correct]
A {\rm LRU-0-Rec} is bad if it contains one or more leakage or regular faults, otherwise it is good. A {\rm LRU-0-Rec} is correct if the {\rm LRU-0-Rec} followed by the ideal {\rm LRU} is equivalent to the ideal {\rm LRU} followed by the ideal {\rm 0-Ga}. (It follows that a good {\rm LRU-0-Rec} is correct, since it contains no faults and so its {\rm LRU}s operate as ideal ones.)
\label{def:gb0}
\end{defi}

These definitions also hold for qubit-preparation and measurement LRU-0-Recs
which do not contain any LRUs. These LRU-0-Recs are good when they are
faultless. This implies that a good measurement LRU-0-Rec can be
replaced by an ideal LRU followed by the corresponding ideal measurement 0-Ga (see the
details in \S \ref{sec:badtofaulty}) allowing us to create ideal LRUs. Similarly, a good qubit-preparation LRU-0-Rec followed by an ideal LRU can be replaced by the corresponding ideal qubit-preparation 0-Ga alone, thus annihilating the ideal LRU.

Consider a fault-tolerant level-$k$ simulation with LRUs appropriately inserted in the circuit as specified above. We can imagine first creating ideal
LRUs out of measurement LRU-0-Recs thereby transforming them to
measurement 0-Ga's. Next, we can imagine propagating these ideal
LRUs to the left through gate LRU-0-Recs thereby transforming them
to gate 0-Ga's. Finally, we can imagine annihilating the ideal LRUs
inside preparation LRU-0-Recs thereby transforming them to
preparation 0-Ga's. Definition \ref{def:gb0} establishes that, if
the corresponding LRU-0-Recs are good, the resulting 0-Ga's will be
the ideal 0-Ga's being simulated. In \S \ref{sec:badtofaulty} we
show that each bad LRU-0-Recs can be transformed to a regular faulty
0-Ga (i.e., one that operates on the system-space).

With the maneuver described above, which we can visualize as an `ideal-LRU wave' propagating
from the right to the left of the level-$k$ circuit, we can transform the level-$k$ simulation
which is subject to leakage faults to an equivalent level-$k$ simulation which is subject to
regular faults alone. In effect, every bad LRU-0-Rec is replaced by a faulty 0-Ga where now
faults act on the system-space. Therefore, we can prove the following lemma.

%
%
%

\begin{lem}[LRU Reduction]
\label{lem:lrureduction} Consider a fault-tolerant level-$k$ simulation $\mathcal{C}$ of some
ideal computation and let $\mathcal{C}_{\rm LRU}$ denote the same simulation where now {\rm LRU}s have been placed preceding every elementary gate {\rm 0-Ga} in $\mathcal{C}$. Let $\mathcal{C}_{\rm LRU}$ be subjected to both leakage as well as regular noise. Then, for any pattern of bad {\rm LRU-0-Recs} in $\mathcal{C}_{\rm LRU}$ corresponding to a set of {\rm 0-Ga}'s $\{\alpha\}$ in $\mathcal{C}$, there exists a mapping to regular noise acting on $\mathcal{C}$ which produces the same computation with regular faults occurring at {\rm 0-Ga}'s in the set $\{\alpha\}$ alone.
\end{lem}

In estimating the threshold for fault-tolerance against leakage
noise we will need to bound the norm of all the operators associated
with bad fault patterns. Lemma \ref{lem:lrureduction} says that
these fault-paths are precisely those that are also bad in the
regular noise model. The only difference is that the 0-Ga's derived
from the LRU-0-Rec by the mapping in Lemma 2 are composite objects with higher error
amplitudes. For the simplest noise model with non-interacting
leakage spaces, the error operator $E$ associated with the LRU-0-Rec
can be bounded as $||E|| \leq \alpha \epsilon$ where $\alpha$ is the
maximum number of location in a LRU-0-Rec. For the more general
noise model we can slightly modify the analysis in \S 11.2 in
\cite{Aliferis05b} in order to start at the LRU-level. The effect is
again that the basic error amplitude $\epsilon$ gets modified to
$\alpha \epsilon$ which has the effect of reducing the threshold by
a factor $1/\alpha$. Thus combining Lemma 2 with the threshold
theorem for regular local noise proves the threshold theorem for the
noise model in \S 2 which we formally state as:

\begin{theo}[Threshold Theorem for Local Leakage Faults]
\label{theo:thresh}
Consider the fault-tolerant simulation including {\rm LRU}s of some ideal computation which is subject to local noise as in \S 2 with strength $\epsilon$ and let $\epsilon$ be smaller than the threshold error strength $\epsilon_{c} \equiv (e A \alpha)^{-1}$ where $\alpha$ is the maximum number
of locations inside a {\rm LRU-0-Rec} and $A$ is the
maximum number of pairs of {\rm LRU-0-Recs} inside any {\rm
LRU-1-exRec}. Let $s$ be the maximum number of locations and let $d$ be the
maximum depth of our {\rm $1$-Recs}.
Then, for any fixed accuracy $\delta$, any ideal computation of size $S$
and depth $D$ can be simulated by such noisy fault-tolerant circuit of size $\it{O}(S(logS)^{log_2 s})$ and depth $\it{O}(D(logS)^{log_2 d})$.
\end{theo}

The value of the constant $\alpha$ appearing in
Theorem$\,$\ref{theo:thresh} depends on our method for realizing
LRUs. If $r$ is the number of elementary operations in a LRU, then
$\alpha$ is at most $2r+1$, since LRU-0-Recs corresponding to
two-qubit gates contain two LRUs. For example, if leakage reduction
is achieved via teleportation then $r=6$ (two qubit-preparations and
a {\sc cnot} to create the Bell state, plus one {\sc cnot} and two
measurements to implement a measurement in the Bell basis). This
implies that the threshold is diminished by at most a factor of $13$
(assuming equal error rates for all locations) in comparison to the
noise threshold for regular errors. For example, for [[7,1,3]] the
threshold {\em lower bound} for a probabilistic model \footnote{For
a coherent non-Markovian error-model this lower-bound has not been
rigorously established in \cite{Aliferis05b}.} $\mathbb{P}_c \geq
2.73 \times 10^{-5}$ would be modified to a lower-bound of
$\mathbb{P}_c \geq 2.1 \times 10^{-6}$ which is probably too
pessimistic. In \S \ref{sec:fewerLRU}, we will consider methods for
potentially improving this threshold bound by omitting LRUs while
still maintaining fault-tolerance.


\subsection{Converting Bad LRU-0-Recs to Faulty 0-Ga's}
\label{sec:badtofaulty}

One technical point in the proof of Theorem \ref{theo:thresh}
concerns moving ideal LRUs to the left of bad LRU-$0$-Recs, thus
transforming them to faulty 0-Ga's. Similar as in the analysis for
moving $k$-decoders past bad $k$-exRecs in
Ref.$\,$\cite{Aliferis05b} we need to use ideal LRUs that are
invertible operations. Both ideal LRUs as well as ideal decoders
generate a syndrome. If the ideal LRU is perfect teleportation, the
syndrome consists of Bell measurement bits. One can define a
coherent invertible teleportation in which one rotates to the Bell
basis and performs controlled $\sigma_{\rm x}$ and $\sigma_{\rm z}$
operations on the outgoing qubit. Furthermore, it is possible to
{\em define} the action of the controlled $\sigma_{\rm x}$ and
controlled $\sigma_{\rm z}$ operation on the extended (leaky) input
space such that the target qubit only has support on the
system-space. Therefore, even for leaky inputs, the output qubit of
the ideal coherent LRU has no leakage fault (leakage will be
confined to the qubits carrying the syndrome information).

Let us denote an ideal LRU discarding its syndrome as ${\cal LRU}$
and the coherent version of an ideal LRU as $\mathfrak{LRU}$. First,
consider those qubits inside the level-$k$ simulation that are
measured. A measurement can be viewed as an ideal LRU followed by
another measurement on the qubit output from the LRU. In
mathematical terms, before every measurement 0-Ga ${\cal M}$ we can
insert an ideal $\mathfrak{LRU} \ast \mathfrak{LRU}^{-1}$ and
identify ${\cal M}'=\mathfrak{LRU}^{-1} \ast {\cal M}$ as a
0-measurement which acts on the system-space (and the syndrome
space). As we will explain below, we can propagate these
$\mathfrak{LRU}$s created out of measurements 0-Ga's to the left
through good LRU-0-Recs transforming them to the ideal 0-Ga's they
contain just as we would propagate ideal LRUs which discard their
syndrome. When we encounter a bad LRU-0-Rec, we proceed by inserting
a resolution of the identity in the form of an ideal (coherent) LRU
and its inverse, $I= \mathfrak{LRU} \ast \mathfrak{LRU}^{-1}$,
preceding it in all its inputs. The leading $\mathfrak{LRU}$(s) in
the $\mathfrak{LRU} \ast \mathfrak{LRU}^{-1}$ pair(s) can now be
moved further to the left through the remaining LRU-0-Recs, until we
encounter another bad LRU-0-Rec and repeat the same trick.

It remains to justify why a bad LRU-0-Rec grouped together with the
$\mathfrak{LRU}^{-1}$(s) preceding and the $\mathfrak{LRU}$(s) (or
${\cal LRU}$(s)) succeeding it can be interpreted as a faulty 0-Ga.
In the combined operation
\beq \mathfrak{LRU}^{-1} \ast {\rm LRU} \ast \mbox{0-Ga} \ast \,
\mathfrak{LRU} \, \, , \label{eq:lrugate1} \eeq
the $\mathfrak{LRU}^{-1}$ to the left has an input in the
system-space and also some input syndrome coming from the
$\mathfrak{LRU}$ pairing with it. Similarly, the
$\mathfrak{LRU}$ on the right outputs a state in the system-space of
the corresponding qubit (since ideal) and some syndrome. For a good LRU-$0$-Rec
(i.e., one with no faults), this operation would be equivalent to applying
the ideal 0-Ga on the input of the $\mathfrak{LRU}^{-1}$ 
in addition to acting on the syndrome. We have already stated this
property as the correctness of good LRU-0-Recs, for the case when
ideal LRUs discard their syndrome. However, note that, even with
ideal LRUs retaining their syndrome, the action on the system-space
of the qubit and the syndrome has to be uncorrelated, otherwise
discarding the syndrome (e.g., the Bell measurement outcomes in
teleportation) would have destroyed the coherence of the teleported
input contrary to our correctness property. For a bad LRU-$0$-Rec instead,
this operation can be viewed as some 0-Ga acting on the input qubit of the $\mathfrak{LRU}^{-1}$. Since this 0-Ga will not necessarily be equal to the ideal operation that is simulated by the LRU-$0$-Rec, it corresponds to a regular faulty 0-Ga. Furthermore, faulty 0-Ga's corresponding to different bad LRU-$0$-Recs may be {\em correlated} as they share access to the syndrome generated by the ideal coherent LRUs.

\section{Limiting The Use of LRUs}
\label{sec:fewerLRU}

In the previous section we assumed that LRUs were inserted on all
inputs of every elementary gate in our fault-tolerant simulation. We
did this in order to be able to interpret every good LRU-0-Rec as an
ideal 0-Ga irrespective of whether its input has leaked or not. In
this section we would like to find out what happens when we omit
LRUs between successive 0-Ga's.

It is clear that it can be advantageous to omit LRUs since the error
rate of a memory location is typically less than the error rate of a
LRU. In what follows we will give conditions for when this omission
is allowed and show how Theorem $\,$\ref{theo:thresh} can be proved
in those cases.

We would like to note that the idea of omitting LRUs and the
modified threshold analysis that we will present is also of interest
when dealing with regular errors. In that case it may be similarly
advantageous to omit error-correction between the application of
successive encoded gates. In particular, for low memory error rates
it may be advantageous to replace a wait location by a sequence of
wait locations at the next level of concatenation instead of a wait
location on the block followed by error-correction on the block.

%
%

Let us begin by defining {\em stretched} LRU 0-rectangles.
\begin{defi}[Stretched LRU 0-Rectangles]
A stretched {\rm LRU-0-Rec} ($\,${\rm LRU-0-StrRec}) is the union of two or more consecutive {\rm LRU-0-Recs} with any {\rm LRUs} between {\rm 0-Ga's} omitted.
\end{defi}

An example of a {\em stretched} LRU-0-Rec is shown in the Fig.$\,$2.
If this circuit contains no faults, then the LRUs on the inputs will
operate ideally and the stretched LRU-0-Rec will be correct. In this
case omitting the intermediate LRU has no consequence. However,
assume a leakage fault only occurred in the first 0-Ga. Then we
cannot interpret the second 0-Ga as an ideal 0-Ga although it is
executed without fault. This is because if the input to that 0-Ga is
not in the system-space of the qubit, then its action can be in
principle arbitrary in the extended space of all its inputs. Thus we
have to assume that if the first 0-Ga fails the second 0-Ga will
fail as well.
\begin{figure}[h]
\begin{center}
\vspace{0.2cm} \epsfig{file=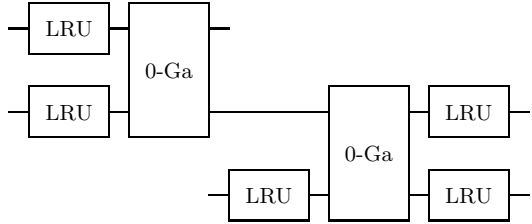,width=7.0cm}
\label{StRec-example} \caption{\footnotesize{An example of a
LRU-0-StrRec obtained by stretching two LRU-0-Recs.}}
\end{center}
\end{figure}

This example shows that LRU-0-StrRecs will be allowed in our
circuits if our fault-tolerant design guarantees that arbitrary
failures at \emph{all} 0-Ga's contained in it will not lead to an
encoded error and a crash.  For instance, a case where LRUs can be
omitted is a sequence of single qubit gates and/or memory locations
on a single qubit: instead of putting LRUs between all locations it
suffices to put a LRU at the beginning of the sequence. The reason
is that a sequence of failed single-qubit locations is no worse than
just one of these locations failing, since it can cause an error in
at most one qubit within an encoded block.

The definition of goodness for LRU-0-StrRecs is the same as for
LRU-0-Recs given in Definition$\,$\ref{def:gb0} (i.e., we will say
that an LRU-0-StrRec is good if it contains no faults, and bad
otherwise). In order to achieve fault-tolerance, we need to be safe
that a single bad LRU-0-StrRec inside an LRU-1-exRec still makes,
after the LRU-Reduction Lemma 2 is applied, for a $1$-exRec which is
correct in the sense defined in Ref.$\,$\cite{Aliferis05b}. As
already discussed, this will not be true for arbitrarily stretched
rectangles, since a leakage fault in a single location can cause the
failure of {\em all} subsequent locations inside the same
LRU-0-StrRec.

%
%

Ref.$\,$\cite{Aliferis05b} lists a set of properties that $1$-EC and
$1$-Ga's must satisfy in order to achieve fault-tolerance against
regular faults. Here, we generalize these properties to
LRU-1-gadgets which use (stretched) LRU-0-Recs. For simplicity we only state the properties for fault-tolerant simulations protected by concatenated distance-3 codes.

\vspace{0.2cm}
\noindent {\it {\bf Properties of LRU-1-gadgets for distance-3 codes}} \vspace{0.1cm} $\\$
\noindent 0$'$,$\,$0. {\em If a {\rm LRU-$1$-EC} contains exactly one bad {\rm LRU-0-(Str)Rec}, 
then it takes an arbitrary input to an output which deviates by at
most a weight-one regular or leakage error-operator from the
code-space, and an input without errors to an output with at most one
leakage or one regular error.}

\noindent 1,$\,$2. {\em If a {\rm LRU-$1$-EC} contains no bad {\rm LRU-0-(Str)Recs}, 
then it takes any input to an output in the code-space, and an input with at most one leakage or one regular error to an output with no errors.}

\noindent 3. {\em If a {\rm LRU-$1$-Ga} contains no bad {\rm LRU-0-(Str)Recs}, 
then it takes an input with at most one leakage or one regular error
in all blocks to an output with at most one leakage or one regular
error in each block.}

\noindent 3$'$. {\em If a {\rm LRU-$1$-measurement} contains no bad
{\rm LRU-0-(Str)Recs}, then it produces the ideal measurement
outcome if its input has at most one leakage or one regular error.}

\noindent 4. {\em If a {\rm LRU-$1$-Ga} contains exactly one bad {\rm LRU-0-(Str)Recs}, 
then it takes an input without errors to an output with at most one
leakage or one regular error in each block.}

\noindent 4$'$. {\em If a {\rm LRU-$1$-preparation} contains exactly
one bad {\rm LRU-0-(Str)Recs}, then its output has at most one
leakage or one regular error. If a {\rm LRU-$1$-measurement}
contains exactly one bad {\rm LRU-0-(Str)Recs}, then it produces the
ideal measurement outcome if its input has no errors. }

It is straightforward to check that these properties are sufficient
for proving that a LRU-1-exRec containing a single bad stretched
LRU-0-Rec will result in a correct equivalent 1-exRec after Lemma 2
is applied. Therefore, if stretching is done such that these
properties are satisfied, Theorem 1 can be proved as before. Let us
now then consider the possible effects of stretching on the {\em
value} of the threshold.

One way of looking at stretching LRU-0-Recs is through the
notion of benign and malignant faults introduced in Ref.$\,$\cite{Aliferis05b}. A single fault inside a stretched
LRU-0-Rec can cause all subsequent 0-Ga's inside it to fail and will thus correspond
to multiple faulty 0-Ga's after applying Lemma 2. If these faults are benign, they will not cause
an encoded error on the data and one is allowed to stretch the rectangle while
maintaining fault-tolerance.

The use of LRU-0-StrRecs necessitates a new analysis of which sets
of leakage or regular faults are benign in the LRU-1-exRecs (no
modification would be needed for determining malignancy at levels
higher than the first). For example, it could be that for regular faults two
particular 0-Ga's form a benign pair. Let us assume that each of
these 0-Ga's is the first 0-Ga in a LRU-0-StrRec containing each,
say, two 0-Ga's as in Fig.$\,$2. A leakage fault occurring in these
0-Ga's also causes failure of the next 0-Ga that is part of the
LRU-0-StrRec. Reinterpreted as regular faults, we now have four
regular faults, which may not form a benign fault-pattern.

If benign fault-patterns for regular errors stay benign
fault-patterns for leakage faults, we could say that our stretching
is {\em benign}. Whether or not stretching is benign can only be
found out by carefully considering any particular stretched level-1 LRU-circuit. In this paper we will not embark on such an
analysis, even though we believe that most but not all stretching is benign. Certainly, stretching in transversal parts of LRU-1-exRecs is benign and the situation only becomes complex inside the non-transversal
ancilla-preparation procedures used for error-correction. If we
refrain from carrying out this more detailed analysis, we can only
establish a threshold lower bound by simply counting all pairs of
locations in a (worst-case) LRU-1-exRec. For [[7,1,3]] and the
stretched LRU-circuits given in Appendix A, this gives a lower bound
of $\mathbb{P}_c \geq {{1247 \choose 2}}^{-1} \approx 1.28 \times
10^{-6}$ \footnote{Our {\sc cnot} 1-exRec contains 575 locations as
in Ref. \cite{Aliferis05b}. In the {\sc cnot} LRU-1-exRec, LRUs need
only be placed inside the four error-corrections as shown in
Appendix \ref{sec:example}. Each of our LRU-1-ECs contains
$4\times7=28$ LRUs, and in turn each LRU contains $6$ locations if
it is implemented by teleportation.}. We believe however that the
real threshold will be closer to the threshold for regular noise.

\section{Leakage in Measurement-Based Computation}
\label{sec:meas-based}

Since not only quantum states but also quantum gates can be
teleported \cite{Gottesman99} it is intriguing to consider combining
gate implementation with teleportation in a single computational
step. An embodiment of this idea is found in measurement-based
models of computation. We will focus on one such model, the graph-state model
\cite{Raussen03} and discuss computation using graph states of a
specific `standard' form. Graph states of this form are of potential
practical interest as they correspond to states that can be prepared
in several proposals relating to quantum computation with
non-deterministic gates (e.g.,\cite{Nielsen04, Browne04,Lim05}).

We consider {\em quantum circuit simulations} by the graph-state
model that proceed by preparing the appropriate graph state and
simulating each quantum gate by executing single-qubit measurements
in the appropriate time-ordering and bases (these bases can change
dynamically and depend on previous outcomes during the simulation).
In the presence of (leakage) noise the corresponding fault-tolerant
circuits can be simulated instead. A threshold theorem for regular
local (non-Markovian) noise has been proved for these fault-tolerant
simulations \cite{Nielsen04b, Aliferis05a} which is analogous to the
threshold theorem for computation by quantum circuits. Our goal in
this section is to supplement these results and show that leakage
reduction is automatically achieved when the graph state is of a
specific form.

To specify this `standard' form, we consider building the graph
state out of basic units that simulate the two-qubit gate $U=\,
\left (U_x(\phi_1)U_z(\theta_1) \otimes U_x(\phi_2)U_z(\theta_2)
\right)\,${\sc cphase}. Here, $U_x(\phi)$ (resp. $U_z(\theta)$) is a
single-qubit rotation around the $x$-axis (resp. $z$-axis) by an
angle $\phi$ (resp. $\theta$), and {\sc cphase} acts in the
computation basis as {\sc cphase}$|i\rangle \otimes |j\rangle =
(-1)^{ij} |i\rangle \otimes |j\rangle$; $i,j\in \{0,1 \}$. Clearly,
any quantum computation can be efficiently realized by a sequence of
these universal gates. This basic unit is shown in Fig.$\,$3(a). In
Fig.$\,$3(b) we give an example of the composition of two units to
simulate a sequence of two 0-Ga's such as the ones in Fig.$\,$2.

\begin{figure}[h]
\begin{center}
\begin{tabular}{ccc}
\hspace{0.6cm} (a) & \hspace{0.5cm} & (b) \vspace{0.3cm} \\
\hspace{1cm} \parbox{3.2cm}{\epsfig{file=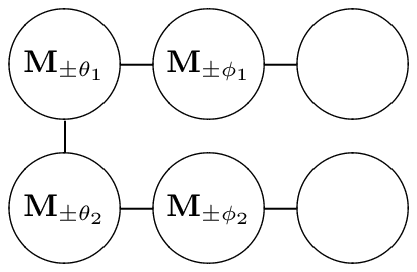,width=3.2cm}} & &
\parbox{6.9cm}{\epsfig{file=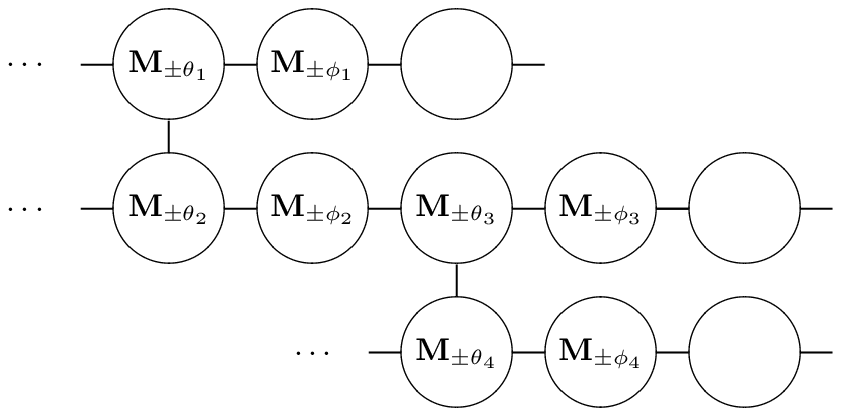,width=6.9cm}}
\end{tabular}
\vspace{0.1cm} \label{fig:meas-based} \caption{\footnotesize{Circles
represent qubits initialized in the state $|+\rangle$. Edges denote
{\sc cphase} gates, and M$_{\omega}$ denotes a measurement of the
observable $cos(\omega)\sigma_{\rm x} + sin(\omega)\sigma_{\rm y}$.
(a) The basic unit in the graph-state `standard' form which
simulates the gate $\left (U_x(\phi_1)U_z(\theta_1) \otimes
U_x(\phi_2)U_z(\theta_2) \right)\,${\sc cphase}. The $\pm$ signs on
the measurement angles are determined by the measurement outcomes
obtained at previous times. (b) The composition of two basic units
to simulate a sequence of two two-qubit gates.}}
\end{center}
\end{figure}

The basic unit of Fig.$\,$3(a) can now be viewed as a new type of
LRU-0-Rec. Indeed, if its two input qubits are in the system-space,
the computation realized by a basic unit is formally equivalent to
applying a {\sc cphase} gate {\em followed} by the teleportation of
single-qubit gates on each qubit (see e.g.,
\cite{Verstraete03,Aliferis04,Childs04}). Let us call each such
sub-pattern realizing a single-qubit gate teleportation a 0-Ga-LRU,
where the 0-Ga should be understood to be the gate being teleported.
0-Ga-LRUs do in fact act as LRUs since, even if one of their input
qubits is in the leakage-space, fresh qubits in the system-space are
always produced at the output as a result of teleportation.

This observation shows that we should consider each basic unit as
realizing a LRU-0-Rec where the 0-Ga (i.e., a {\sc cphase} gate
here) is {\em succeeded} by 0-Ga-LRUs. Moreover, if this LRU-0-Rec
contains no faults, it realizes the ideal operation if its input is
in the system-space. In order to give conditions for this to be the
case, we will need to define LRU extended rectangles (LRU-0-exRecs)
that include an LRU-0-Rec grouped together with its {\em preceding}
0-Ga-LRUs. We can therefore modify Definition$\,$\ref{def:lru-rect}
as follows (see Fig.$\,$1).
\begin{defi}[LRU-0-Recs (revised) and LRU-0-exRecs]
A {\rm LRU-0-Rec} is a {\rm 0-Ga} succeeded by {\rm 0-Ga-LRUs} on
all its outputs. A {\rm LRU-0-exRec} is a {\rm LRU-0-Rec} preceded
by {\rm 0-Ga-LRUs} on all its inputs. \label{def:lru-rect2}
\end{defi}
\begin{figure}[h]
\begin{center}
\epsfig{file=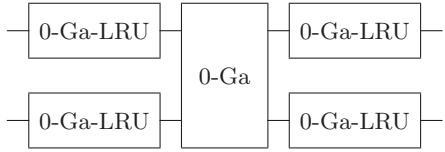,width=5.9cm} \label{lru-exRec}
\caption{\footnotesize{An example of an LRU-0-exRec corresponding to
a two-qubit 0-Ga.}}
\end{center}
\end{figure}

Similar to Definition$\,$\ref{def:gb0}, we can now define
LRU-0-exRecs to be {\em good} when they contain no faults. Then, it
is easy to see that the LRU-0-Rec contained in a good LRU-0-exRec is
{\em correct}, in the sense that applying the LRU-0-Rec and then
passing the outputs through ideal LRUs is the {\em same} as applying
the ideal LRUs first, succeeded by the ideal 0-Ga that the LRU-0-Rec
simulates. Here are the revised definitions of goodness and
correctness at level 0.
\begin{defi}[LRU Good, Bad, Correct for LRU-0-exRecs]
A {\rm LRU-0-exRec} is bad if it contains one or more leakage or
regular faults, otherwise it is good. Two bad {\rm LRU-0-exRecs} are
independent if they are nonoverlapping or if they overlap and the
earlier {\rm LRU-0-exRec} is still bad when the shared {\rm
0-Ga-LRUs} are removed. A {\rm LRU-0-Rec} is correct if the {\rm
LRU-0-Rec} followed by ideal {\rm LRU}s is equivalent to the ideal
{\rm LRU}s followed by the ideal {\rm 0-Ga} it simulates. (It
follows that the {\rm LRU-0-Rec} contained in a good {\rm
LRU-0-exRec} is correct, since the good {\rm LRU-0-exRec} contains
no faults and so its leading {\rm LRUs} operate ideally.)
\label{def:gb0-mod}
\end{defi}

The notion of independence is needed to guarantee that for two
successive (and therefore overlapping) LRU-0-exRecs to be bad there
need to be at least two faults.

With these new definitions we can now prove a lemma analogous to
Lemma 2 in a similar way as in \S \ref{sec:leakanalysis}. The proof
will again proceed via the LRU-wave maneuver. As the ideal LRUs
march to the left, they map every LRU-0-Rec contained inside a good
LRU-0-exRec to the ideal 0-Ga that the LRU-0-Rec simulated and they
transform every LRU-0-Rec inside a bad LRU-0-exRec into some faulty
0-Ga. These faulty 0-Ga's can be correlated as they share access to
both the syndrome that is generated by our ideal LRUs and to the
actual syndrome information that is generated by the measurement
outcomes of the graph-state simulation itself. Formally treating the
case of bad LRU-0-exRecs can be done in the same way as in
\S\ref{sec:badtofaulty} with the difference that now $\mathfrak{LRU}
\ast \mathfrak{LRU}^{-1}$ pairs are inserted preceding the {\rm
entire} bad LRU-0-exRec and not just the LRU-0-Rec contained in it.
This will result in the truncation of one of the trailing 0-Ga-LRUs
of the preceding LRU-0-exRec. Provided this truncated LRU-0-exRec is
good, we can next move the ideal LRUs to the left thereby
transforming the (truncated) LRU-0-Rec to the ideal 0-Ga that it
simulates \footnote{Moving ideal LRUs through good truncated
LRU-0-exRecs is analogous to moving ideal $k$-decoders through good
truncated $k$-exRecs (see \cite{Aliferis05b}) and can be
accomplished in essentially the same way.}. Because the LRU-0-Rec is
truncated, this ideal 0-Ga differs from the ideal 0-Ga that the full
LRU-0-Rec simulates by the gate being teleported in the truncated
0-Ga-LRU. However, we can insert these single qubit rotations and
their inverses and absorb the inverses in the faulty 0-Ga that
replaced the bad LRU-0-exRec. Therefore the entire LRU-wave maneuver
can be completed without change.

Consider now applying the LRU-wave maneuver described above to a
fault-tolerant circuit simulation in the graph-state model subject
to our leakage noise model. In the resulting equivalent simulation,
all LRU-0-Recs contained in bad LRU-0-exRecs are mapped to regular
faulty 0-Ga's while LRU-0-Recs contained in good LRU-0-exRecs are
mapped to the ideal 0-Ga's that the LRU-0-Recs simulate. For
LRU-0-Recs containing regular faults (or no faults) this mapping is
justified by the proofs given in Refs.$\,$\cite{Nielsen04} and
\cite{Aliferis05a} and we will not give the full details here. Thus,
using the LRU-wave maneuver we can map the problem of
fault-tolerance in a graph-state simulation subject to local
(leakage) noise to the problem of fault-tolerance in the circuit
model under local regular noise for which the proof in
Ref.$\,$\cite{Aliferis05b} applies. This proves the existence of an
accuracy threshold for fault-tolerant circuit simulations using
graph states in the standard form of Fig.$\,$3 in the mixed local
error model specified in \S\ref{sec:leakmodel}.

\section{Discussion}
\label{sect:discuss}

Our fault-tolerance analysis has been based on LRUs implementable by
quantum teleportation. In various experimental schemes it may also
be possible to detect leakage faults. For instance, in optical
quantum computation, parity-measurements of the photon occupation
number in different modes can be performed which will indicate
whether photon loss has occurred \cite{Knill00}. Another example is
the detection of photons from a cavity with trapped ions
\cite{Pellizzari95}. Some leakage detection is also present in the
solid-state scheme where a qubit is encoded in a two electron-spin
state of a double quantum dot \cite{Petta05}. Such leakage detection
can be used in the following ways.

Leakage detection acting on single or multiple qubits will tell us
whether leakage occurred but not necessarily which qubit was
affected. If leakage detection is used on ancillae, we can throw away the leaked qubits and repeat the ancilla preparation circuit a fixed number of times. If leakage detection is used on data qubits,
we can follow by applying teleportations on all qubits that
potentially leaked. Or, if leakage detection specifically points to
one qubit, we can replace the qubit simply by a fresh physical
qubit. Otherwise, if no leakage is detected, we can omit
teleportation or qubit replacement. Leakage detection may of course fail, i.e. qubits can leak
but this leakage can go undetected. This case of unsuccessful
leakage detection is similar to having a leaky LRU. Thus leakage detection
performed before each gate or continuously during gates provides a
fault-tolerant way of dealing with leakage.

More accurately, if leakage detection per qubit is possible, we can
define a Leakage Detection Unit or LDU. The LDU acts as a LRU and in
addition gives us a bit that says whether leakage occurred or not.
So the idea is to replace LRUs by LDUs inside the LRU-0-Recs or
LRU-0-StrRecs. Since all single leakage-fault events will thus be detected, and because faults in the LDUs
themselves can either incorrectly signal leakage detections or at
worst cause the LRU-0-Rec in which they are contained to fail, Definition$\,$\ref{def:gb0} still applies and
Lemma 2 can be proved without modification.

It is clear that leakage reduction implemented by teleportation may require frequent measurement.
This could potentially be problematic in situations where measurements are slow compared to fundamental gate-times, as e.g. in many (solid-state) implementations of quantum computation. However, data qubits do not always
need to wait for the measurements to finish. As with regular error-correction where measurements are used, the
Pauli corrections that result from teleportation can in many cases
be kept in a classical memory and they need only be used to adapt
the non-Clifford parts of the computation. In fact, since
non-Clifford gates are used in logical parts of the computation and
not during error-correction, the measurement outcomes of
teleportations can be combined with the results of error-correction
to give a joint Pauli correction operator. Therefore, assuming
classical computation is fast and robust, leakage reduction via
teleportation will not suffer any additional time overheads than the
overheads already associated with regular error-correction that uses
measurements.

\section{Acknowledgements}
We are grateful to Debbie Leung for discussions. PA would like to thank the IBM
Quantum Information Group for its hospitality and acknowledges
support by the Canadian Institute for Advanced Research and by US
NSF under grant no.$\,$PHY-0456720. BMT acknowledges support by the
NSA and the ARDA through ARO contract number W911NF-04-C-0098. Our quantum circuit diagrams were drawn using the Q-circuit
\LaTeX\ macro package by Steve Flammia and Bryan Eastin.



\newcommand{\etalchar}[1]{$^{#1}$}


\appendix


\section{Leakage-resistant circuits for the [[7,1,3]]}
\label{sec:example}

In this section, we consider 1-EC circuits associated with the
Steane [[7,1,3]] code in order to illustrate the idea of stretching
LRU-0-Recs. Our discussion can in fact be applied to
error-correction circuits for a much wider variety of CSS codes but
we will not give such general exposition here.

Fig.$\,$5(a) shows the $1$-EC circuit diagram with error-correction
performed according to Steane's method \cite{Steane96b}. In
Fig.$\,$5(b), LRUs are placed in this circuit to deal with leakage
faults. The resulting LRU-1-EC contains eleven LRU-0-StrRecs in
total: Each of the four ancilla-encoding circuits constitutes a
separate LRU-0-StrRec (containing no LRUs), and in addition there
are seven LRU-0-StrRecs containing the succeeding transversal
operations acting in parallel.

\begin{figure}[h]
\begin{center}
\begin{tabular}{ccc}
(a) & \hspace{1.5cm} & (b) \vspace{0.1cm} \\
\epsfig{file=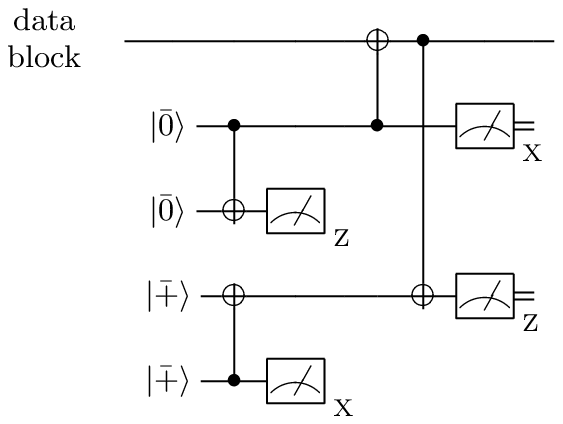,width=4.75cm} & & \epsfig{file=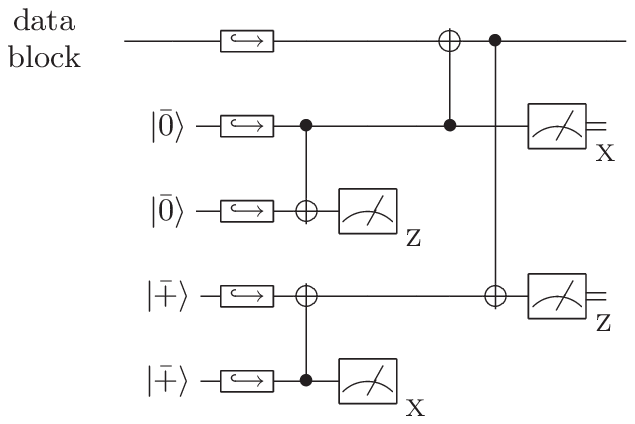,width=5.45cm}
\end{tabular}
\label{Steane-EC} \caption{\footnotesize{(a) A 1-EC circuit. Ancilla verification is done by comparing two identical logical states and rejecting whenever an error is detected. {\sc cnot} gates and measurements are realized transversally. (b) The corresponding LRU-1-EC circuit. Transversal LRUs are indicated by `hook' arrows. The encoding circuits for the logical states $|\bar{0}\rangle$ and $|\bar{+}\rangle$ contain no LRUs and are suppressed (e.g., see \cite{Aliferis05b}).
}}
\end{center}
\end{figure}
The circuit in Fig.$\,$5(b) satisfies the properties in \S
\ref{sec:fewerLRU} due to the ancilla verification method we have
chosen (e.g., see \S 7.2.1 in Ref.$\,$\cite{Aliferis05b} for a
recent review). This method consists in encoding two identical logical
ancillae (i.e., two $|\bar{0}\rangle$ states or two
$|\bar{+}\rangle$ states) and checking suitable parities before
connecting the verified ancilla with the data. Because verification
succeeds as long as one of the ancilla-encoding circuits contains no
faults independently from the number of faults in the other encoder,
it is sufficient in the presence of leakage faults to ensure that
the state that comes out of the two encoders is in the system-space.
This is achieved by placing LRUs immediately after the ancilla-encoding
circuits. LRUs are also placed in the data block before it is
connected to the ancillae in order to convert existing leakage
errors to regular errors before error-correction is performed.

An alternative $1$-EC circuit due to Knill \cite{Knill04}
is shown in Fig.$\,$6(a). In Fig.$\,$6(b) LRUs are placed to deal
with leakage faults resulting in eleven LRU-0-StrRecs similarly to
Fig.$\,$5(b). This `teleported error-correction' method is
especially advantageous against leakage as it includes the
teleportation of the logical state in the input code block in a natural way: no LRUs need to
be placed in the input block since the encoded input state is
naturally teleported to a fresh block at the output. However LRUs
still need to be inserted immediately after the ancilla-encoding
circuits just as in Fig.$\,$5(b).

\begin{figure}[t]
\begin{center}
\begin{tabular}{ccc}
(a) & \hspace{1.5cm} & (b) \vspace{0.1cm} \\
\epsfig{file=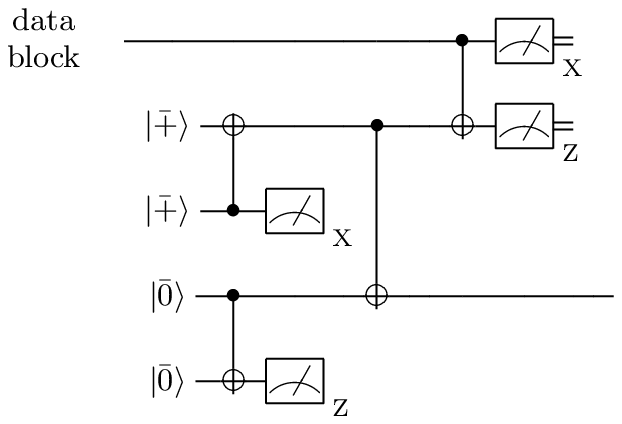,width=5.2cm} & & \epsfig{file=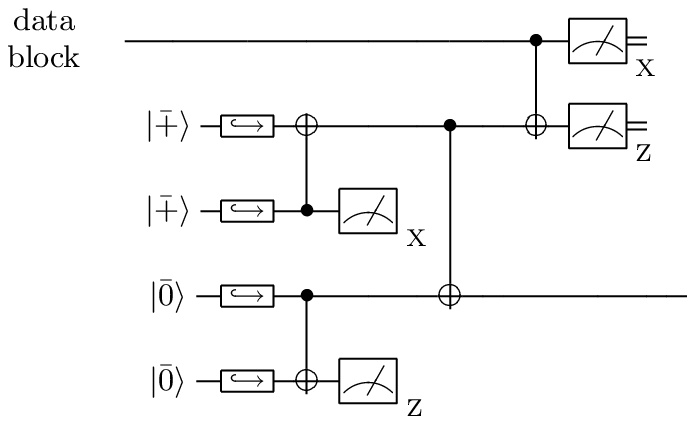,width=6.0cm}
\end{tabular}
\label{ECT} \caption{\footnotesize{(a) An alternative 1-EC circuit. A logical Bell-state is prepared using verified
logical $|\bar{0}\rangle$ and $|\bar{+}\rangle$ ancillae. Transversal measurements in the Bell basis are then performed
between the input data block and one half of the logical Bell-state. (b) The corresponding LRU-$1$-EC circuit. Note that no LRUs are required on the input data block.}}
\end{center}
\end{figure}


\end{document}